\documentclass{JINST}
\def\centeron#1#2{{\setbox0=\hbox{#1}\setbox1=\hbox{#2}\ifdim
\wd1>\wd0\kern.5\wd1\kern-.5\wd0\fi
\copy0\kern-.5\wd0\kern-.5\wd1\copy1\ifdim\wd0>\wd1
\kern.5\wd0\kern-.5\wd1\fi}}
\def\centerover#1#2{\centeron{#1}{\setbox0=\hbox{#1}\setbox
1=\hbox{#2}\raise\ht0\hbox{\raise\dp1\hbox{\copy1}}}}
\def\centerunder#1#2{\centeron{#1}{\setbox0=\hbox{#1}\setbox
1=\hbox{#2}\lower\dp0\hbox{\lower\ht1\hbox{\copy1}}}}
\def\lsim{\;\centeron{\raise.35ex\hbox{$<$}}{\lower.65ex\hbox
{$\sim$}}\;}
\def\gsim{\;\centeron{\raise.35ex\hbox{$>$}}{\lower.65ex\hbox
{$\sim$}}\;}
\title{Quartz Cherenkov Counters for Fast Timing: QUARTIC}

  \author{M.G.Albrow$^a$,
  Heejong Kim$^b$, S.Los$^a$,  M.Mazzillo$^c$,E.Ramberg$^a$, 
  A.Ronzhin$^a$, V.Samoylenko$^d$, H.Wenzel$^a$, 
   and A.Zatserklyaniy$^e$\\
  
  \llap{$^a$} Fermi National Accelerator Laboratory, Wilson Road, Batavia, IL 60510, U.S.A.\\
  \llap{$^b$}  Enrico Fermi Institute, University of Chicago, Chicago, IL 60637, USA\\
   \llap{$^c$} STMicroelectronics, Catania 95121, Italy \\
  \llap{$^d$}  Institute for High Energy Physics, Moscow region, RU-142284 Protvino, Russia \\
  \llap{$^e$} Institute for Particle Physics, University of California, Santa Cruz, CA 95064, USA \\
 E-mail: \email{albrow@fnal.gov}}

\abstract{
We have developed particle detectors based on fused silica (quartz) Cherenkov radiators read out 
with microchannel plate photomultipliers (MCP-PMTs) or silicon photomultipliers (SiPMs) for high precision
timing ($\sigma_t \sim$ 10-15 ps). One application is to measure the times of small angle protons 
from exclusive reactions,
$p+p \rightarrow p+X+p$, at the Large Hadron
Collider, LHC. They may also be used to measure directional particle fluxes close to external or stored beams.
The detectors have small areas (cm$^2$), but need to be active very close
($\sim$ 4 mm) to the intense LHC beam, and so must be radiation hard and nearly edgeless. 
We present results of tests of detectors with quartz bars inclined at the Cherenkov angle, and with bars in the form of an
``L" (with a 90$^\circ$ corner). We also describe a possible design for a fast 
timing hodoscope with few mm$^2$ elements.}


\begin{document}


\section{\textbf{Introduction}}

We describe the development of particle detectors with time resolution $\sigma_t ~\sim$ 15 ps that have the following characteristics:
(a) they are edgeless, with active area within about 200 $\mu$m of the outside (b) they are radiation hard enough to be
close to primary 
beams (c) they
have segmentation, allowing independent timing of several particles in a single 1 ns pulse. The main limitation is that they have
small active areas, $\sim$ 4 cm$^2$, which is however adequate for our applications. They are compact, and the best time resolution is achieved by having
multiple measurements, e.g. four or more modules in-line, giving what we call a time-track. This approach allows monitoring of the resolution 
and efficiency of each module, and relaxes the requirements on the electronics compared with having a single measurement.
The detectors use Cherenkov light in quartz (fused silica) radiators, read out by either microchannel plate photomultipliers (MCP-PMTs) or
silicon photomultipliers (SiPMs).

  Detectors measuring the time of particles have been used in high energy physics experiments for decades,
  usually to measure the speed of particles and hence, in conjunction with momentum or (rarely) energy, to determine
  their mass and hence their identity.  Colliding beam experiments typically require large area detectors ($\sim 10$~m$^2$) at a few m
  distance, have time resolutions $\sim$ 100 ps and can separate $\pi$, K and p in the momentum region $\sim$~1
  GeV/c. Detectors with similarly large areas and better time resolution, by an order of magnitude,
  are being actively developed~\cite{frisch}.

A different use of high resolution timing is to determine the particle path length, especially for
  ultrarelativistic particles ($\beta \approx 1.0$). This can be used to determine the position of origin of a pair of particles,
  as in positron emission tomography, PET, where the particles are $\gamma$-rays from $e^+e^-$ annihilation. 
  We report here on the development of detectors with resolution $\sigma_t \sim$ 15 ps
  to determine the position of origin of a pair of high energy protons, if indeed they came from the same collision. That corresponds to only 4.5 mm of light travel time, and
  scintillation counters of large dimensions with conventional photomultipliers (PMTs) are excluded.
  Cherenkov counters have prompt light emission, and MCP-PMTs are inherently
  faster than conventional PMTs, with transit time spreads (sigma), TTS, (for a single photoelectron) or Single Photon Time Resolution, 
  SPTR, of order 30-50 ps. Available SiPMs have SPTR of order 120 ps, similar to conventional photomultipliers, but there is progress in reducing this. The
  intrinsic time resolution due to photoelectron statistics with $N_{pe}$ photoelectrons is then 
  $\sigma_t \sim SPTR/\sqrt{N_{pe}}$, but for a real detector it
  is also subject to geometry, dimensions and electronics.

In this paper we first give a motivation for our studies, followed by some general background information about Cherenkov radiators and photodetectors.
We then discuss three geometries for small quartz Cherenkov counters optimized for fast timimg, followed by some simulations and beam tests. Finally we present a
design for a multi-channel bar array suitable for measuring protons very close to the outgoing LHC beams.

  \subsection{Motivation}
  
  This work was largely motivated by proposals \cite{fp420} 
  to add proton detectors to both ATLAS
  (called AFP for ATLAS Forward Protons) and CMS (called HPS, for High Precision Spectrometers) at 
  the Large Hadron Collider, LHC, to measure ``central exclusive reactions".For a recent review see Ref.\cite{acf}.
  Examples of such exclusive reactions are $p+p \rightarrow p + H + p$ and $p+p \rightarrow
  p + W^+W^- + p$, with no other particles produced. Detecting both protons, and measuring their momenta 
  (after they have traversed LHC magnetic fields),
  enables many properties of the central state, which is detected in the main central detectors, to be determined~\cite{fp420}.
  However the cross sections are expected to be
  very small ($\sim$ 10 fb), requiring high luminosity $L$, e.g. $L = 10^{34}$~cm$^{-2}$~s$^{-1}$, with $\sim$ 25
  interactions per bunch crossing, which has a time spread of $\sigma \sim$ 150 ps, every 25 ns. Then the desired 3-fold coincidence events [pXp] have a large background from
  two-fold pile up [pX][p] or [pp][X], and 3-fold pile up [p][X][p], with obvious notation.
  The time difference $\Delta t$ of 
  the two oppositely directed protons' arrival at far detectors, hundreds of meters down the beam pipes, gives a measure of the 
  collision point, $z_{pp} = \frac{1}{2}.c \Delta t$. A resolution on the time
  difference, $\sigma(\Delta $t$) = \sqrt{2} \times \sigma_t$ of 15 ps would give $\sigma(z_{pp}) = \frac{1}{2} \times 4.5$ mm = 2.25 mm.
   The vertex position $z_X$ of the
  central state [X] can usually be reconstructed with  $\sim 10\;\mathrm{\mu}$m precision, and by matching it to
  $z_{pp}$ a factor $\sim$25 reduction in this pile-up background can be achieved, since the interaction region is much broader
  ($\sigma_z \sim$ 50 mm). In combination with other constraints (e.g. no additional tracks on the central vertex, longitudinal momentum conservation,
  and small transverse momentum, $p_T$, for the central state) this can make exclusive
  Higgs, $W^+W^-$, etc. measurements feasible. Timing resolutions of order $\sigma(\Delta $t$) =$ 10 ps are needed for $L = 10^{34}$ cm$^{-2}$ s$^{-1}$ (assuming 2808
  bunches, i.e. 25 ns bunch spacing).  The time difference measurement between scattered protons was first proposed as
  a means of reducing pile-up in exclusive interactions at the Tevatron~\cite{mmcdf,cdfh}.

  The required detector area is small ($\sim$ 4 cm$^2$).  
  Key requirements, in addition to the time resolution, are \emph{edgelessness} at the level of $\sim$ 100 $\mathrm{\mu}$m on the edge adjacent to the beam, as well as radiation
  hardness to $ \gsim 10^{15}$ protons/cm$^2$~\cite{onel}. In case there is more than one proton in the acceptance from the same bunch crossing, segmentation is also
  required. The protons have been deflected out of the beam by the LHC magnets, but at the
  $z$ position (distance along the beam pipe) of the detectors they are displaced by only a few mm, so any inactive area (on one edge) causes a loss in acceptance.
   We expect a
  proton flux of about $10^{15}$~cm$^{-2}$ per year close to the beam. For both these reasons the photon detector must be
  placed farther from the beam than the radiator; in addition it can then also be shielded from background radiation.
  Replacing the photodetectors on a one-year time scale, if necessary, is feasible; they are accessible and relatively
  inexpensive.
  
  We have tested detectors with various geometries, and here we report on two that can satisfy these stringent requirements.
  Both are called \textsc{quartic} for QUARtz Timing Cherenkov.
   The first (``angled-bar \textsc{quartic}") has quartz bars inclined at the
  Cherenkov angle $\theta_{ch} \sim 48^\circ$ with light detected by MCP-PMTs. 
  The second has the quartz bars in the form of an L (L-bar \textsc{quartic}), one bar being the radiator, R, and the 
  other being the light guide, LG. At the end of the LG bar the light is detected by a SiPM. (It could also be 
  detected by an MCP-PMT, although available devices
  do not have an ideal multi-channel geometry.)
  This novel geometry works because (a) the protons are very nearly parallel to the radiator bar, and (b) the refractive 
  index of quartz is $n(\lambda) >
  \sqrt{2}$. With condition (a) all the Cherenkov light is totally internally reflected (TIR) until it reaches the end 
  of the radiator bar, apart
  from absorption (which is small) and imperfections (on the surface or in the bulk) causing light to be scattered out. 
  With condition (b) the light reflected up the LG bar continues to be totally internally reflected, 
  and is not close to the critical angle. (The angle with respect to the LG bar axis is 90$^\circ - \theta_{ch} \sim
  42^\circ$.) It is important that the bars are not aluminized or 
  wrapped, so that close to 100\% TIR is maintained, and that they are minimally supported (at a few corner points). 
  We describe this more later,
  together with \textsc{geant4} simulations.
  
   Note that the quantity which must be precisely measured is the time difference $\Delta t$ between 
  the protons at the detectors approximately 480~m, and later 840~m, apart.
   This requires a reference time signal (``clock") at each detector with minimal jitter between the left (L) and right (R) detectors,
   $t_L-t_R$. 
  Reference timing systems developed for the International Linear Collider (ILC) have been designed to yield
   an r.m.s. jitter $\sigma_{LR} <$ 1 ps over similar distances, using RF transmission and a phase-locked loop.
  
   A calibration of the absolute time difference (or $z_{pp} = z_X$) can be derived
   from real events of the type $p+p \rightarrow p+X+p$, where $X$ is a set of particles measured in the central
   detector.
   
   While the time difference $t_L-t_R$ gives $z_{pp}$, the absolute time, or $(t_L+t_R)/2$ (minus a constant), would provide another, orthogonal, 
   variable for pile-up
   rejection if the actual event time were known much better than $\sigma_z/c \sim$ 150 ps. The existing ATLAS and CMS detectors do not have such
   capability. It could be made available with large area (and thin) fast timing detectors~\cite{frisch} in a future upgrade, but we do not discuss that here.
   
   A precision track detector, using silicon (or possibly diamond) strips or pixels precedes the timing detectors, giving the
   position and direction of the protons.
    \textsc{quartic} detectors are relatively thick; the nuclear interaction length of quartz is 44 cm
    and the radiation length is 12.3 cm. They should therefore be positioned after the tracking.
    Interactions in the quartz should not degrade the time measurement significantly (and may even improve it!)
   
  Another possible application of these small, edgeless, fast and radiation hard detectors is to measure the fluxes of particles very close to a
  circulating, or external, beam. The detectors we have developed should be suitable for such beam condition monitoring; they are also directional,
  distinguishing ``incoming" and ``outgoing" particles.
  
  Refs~\cite{ronzhinsipm,ronzhin2,ronzhinwfd,ronzhintof,ftdalbrow} report earlier studies by our group 
  on fast timing detectors at Fermilab.
  
  \subsection{Cherenkov detectors for timing}

   Cherenkov light is prompt and therefore ideal for fast timing, although the amount of light is small
compared with scintillator. Radiators should be transparent, i.e. with a long absorption length
$L_{abs}(\lambda)$, where $\lambda$ is the optical wavelength, preferably into the
ultraviolet, $\lambda \approx$ 200 nm, where most photons are generated. While gases, liquids and solids are all possible radiators, the
number of Cherenkov photons radiated is proportional to $1-1/n^2(\lambda)$; more completely
(for charge Q = 1, and $\beta$ = 1):
\[\frac{d^2N}{dxd\lambda}=\frac{2\pi\alpha}{\lambda^2}\left(1-\frac{1}{n^2(\lambda)}\right),\]
where $\alpha$ is the fine structure constant.

The light is emitted along the particle's path in a cone
with half angle (Cherenkov radiation angle) $\theta_{ch}$ given by cos$(\theta_{ch}) = 1/n(\lambda)$.
Solid radiators are much shorter than gases for the same light output, an important
consideration when space is limited. Among solid radiators, fused silica, SiO$_2$, or quartz,
(ultraviolet grade, UVT) is commonly used, and was our choice for these tests. Its refractive index as a function of wavelength is
given in Table 1, together with the light absorption length, and the photon detection efficiency, PDE, of the detectors we used.
The quartz bars we used were supplied by Specialty Glass~\cite{spglass}.

\begin{table}
\begin{center}
\begin{tabular}{|c|c|c|c|c|c|c|}
\hline
\hline
 Wavelength & Refractive & $\theta_{ch}$ & Absorption & PDE (\%) & PDE (\%) \\
    (nm)        &  Index                & (degrees)    &  length (cm)  & MCP-PMT  &  SiPM \\
\hline
  250      &  1.510 & 48.5 & 95 &  20.6 & 0 \\
   300      &  1.488 & 47.8  &  104&  19.1 & 5\\
   350      &  1.475 & 47.3 & 111 &   18.6   & 38 \\
   400      & 1.470  & 47.1 & 120 &  13.8 &  48\\
   450      & 1.465  & 47.0 & 122 & 14.3 &  50 \\
   500      & 1.462  & 46.8 & 125 & 9.4 &   47\\
   550      & 1.460  & 46.8 & 128 & 8.2 &    40\\
   600      &  1.458 & 46.7 & 130 & 7.8 &  30\\
   650      &  1.456 & 46.6 & 130 & 7.3 &  24\\
   700      &  1.455 &  46.6& 130 & 2.0 &  18\\
   750      &   1.450 & 46.4 & 130  & 0.8 &  13 \\
 \hline
 \end{tabular}
 \end{center}
 \caption{Properties of quartz and typical photodetectors in these studies.The MCP-PMT is the Photek PMT210 or PMT240 with fused
 silica windows, The SiPM is the Hamamatsu type S-10362-33 MPPC.}
 \label{props}
\end{table}

The Cherenkov angle in quartz at 250 nm (750 nm) is 48.5$^\circ$ (46.4$^\circ$) respectively, see Table. 1. 
This spans the wavelength range where typical photocathodes are sensitive. We will address
chromaticity later, but for simplicity we use $\theta_{ch}$ = 48$^\circ$ when it is not important.

The Nagoya group~\cite{akatsu}  measured the timing properties of a Cherenkov counter with
 a quartz radiator in-line with a MCP-PMT at the back. If the particles are parallel to the sides
 of the radiator bar, all the Cherenkov light that hits the sides is totally internally reflected, as the Cherenkov angle, $\theta_{ch}$, exactly matches
 the total internal reflection angle (defined with respect to the surface, not the normal)~\footnote{If the radiator bar is clad or immersed
 in a medium with $n > 1$ the light will not be totally reflected.}.

 The approximate rule for the number of photoelectrons in a typical detector is:
   \[N_{pe} \sim 90 \: \mathrm{cm}^{-1} \cdot \mathrm{L(cm)\: sin}^2 \theta_{ch} \sim 50\; \mathrm{cm}^{-1}\times \mathrm{L},\]
   for quartz radiator, which gives 200 photoelectrons for 40 mm. 
   The Nagoya group obtained a time resolution of $\sigma_t$ = 6.2 ps with 3 GeV/c pions ($\beta \sim$ 1).

  \subsection{Photodetectors: Microchannel Plate PMTs} 
  
   Microchannel plate PMTs have a photocathode on a quartz (or similar) window, and photoelectrons
     generate avalanches in thin ceramic plates traversed by holes (pores) with high electric fields (e.g. 2.5 kV per plate, with two
     plates giving a gain of 10$^6$). 
 The single photon time resolution, SPTR, of the MCP-PMT we used
 is about 30 ps, and with $N_{pe}$ photoelectrons one can expect a contribution to the time resolution from photoelectron 
 statistics to be
 $\sigma_t(p.e.) \sim 30/\sqrt{N_{pe}}$ ps. We used a PHOTEK PMT240 directly in the test
 beam as a fast reference time detector in most of our tests; there is enough Cherenkov light generated in the 9 mm quartz window that no additional
 radiator is needed.
 The PMT240 itself cannot be used to detect particles close to a beam as it is not edgeless. (Also the MCP-PMT will 
 not survive the high particle fluxes,
 for reasons of radiation damage and photocathode lifetime.) A potential weakness of MCP-PMTs is that the photocathode
can get damaged by positive ion feedback, which limits their life to typically $10^{14}$ photoelectrons, which may be 
only weeks in the LHC environment. Developments are underway~\cite{jinno} to extend MCP-PMT lifetimes.
 
 \begin{figure*}[t]
\centering
\includegraphics[width=80mm]{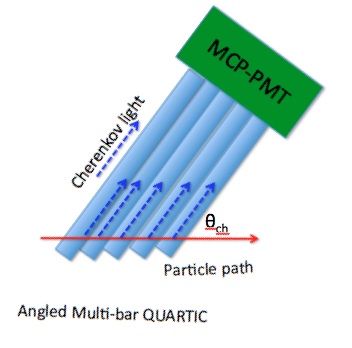}
\caption{Arrangement of an angled-bar \textsc{quartic} with several rows of bars on an MCP-PMT
(single anode or multi-anode).}
\end{figure*}

  \subsection{Photodetectors: Silicon photomultipliers}

     Silicon photomultipliers, SiPMs, are solid state photon counters comprised of a large number of avalanche 
     photodiodes (APDs) or ``pixels" of order 20 $\mu$m
     dimensions, 
     with a high gain (up to $10^6$) in Geiger mode, with an applied voltage just above the breakdown 
     voltage (about 30V to 70V depending on the type).
     Each discharged pixel has a recovery time of order 100 ns, but with e.g. 100 photoelectrons per event and thousands of
     pixels per mm$^2$ this can be acceptable. For the SiPMs the PDE is the product of the quantum efficiency 
     and the fractional area coverage of the APDs. SiPMs are rugged, simple to use and relatively
     cheap \emph{per unit}, but at present are only available
     commercially with effective active areas from 1$\times$1~mm$^2$ to 3.5$\times$3.5~mm$^2$. Smaller SiPMs have less capacitance and so are
     intrinsically faster. The SPTR of typical SiPMs is worse than that of an MCP-PMT, but 
    devices with SPTR $\sim$ 65 ps are becoming available.
  
 \section{Quartz bar geometries}
 
 \subsection{Straight bars, in-line or angled}
   
   We studied three quartz bar geometries: (a) short bars (from 6 mm to 30 mm) 
   in line with the beam (``Nagoya" configuration) (b) long bars (up to 150 mm) inclined at the Cherenkov angle $\theta_{ch} \sim 48^\circ$ 
   (``angled-bar \textsc{quartic}")
   (c) L-shaped bars with a radiator bar (30 mm - 40 mm long) and a light-guide bar at 90$^\circ$ (``L-bar \textsc{quartic}"). Geometry (a) has very efficient light collection, but 
   can only be used when the photodetector can be in the beam. We used it in our tests as a reference counter (in the beam) and to study some general
   properties, e.g. longer radiator bars give more light, but the time spread is also longer; how do these interplay? 
   In this ``in-line" case, if the radiator bar length is $L_{bar}$ the 
      light emitted at the front
      of the bar arrives at the PMT $\Delta t = L_{bar}(n^2 - 1)/c$ later than that emitted at the back (one power of $n$ is from the path
      length and one is from the light speed). For a 40 mm bar this time spread is about 160 ps, so that longer bars, while producing
      more light, may not improve the time resolution. In addition chromatic dispersion should be considered.
      While most of the Cherenkov light is blue/UV, that
light is also slowest and reaches the photodetector later where it is less useful for timing. 
Over 10 cm a 200 nm photon lags behind an 800 nm photon by $L\times c \times \Delta n$ = 100 mm $\times (10ps/3mm) \times [n(\lambda_{200})
-n(\lambda_{800})]$ = 32.7 ps, in addition to any path length differences.
   
   Geometries (b) and (c)
    have the photodetector remote from the beam, which may be essential for mechanical reasons as well as to be in a low radiation environment. For geometry (b)
    the radiator bar is long and rotated by an angle so that the photodetector is remote from the beam. A special configuration is
    $\theta_{bar} = \theta_{ch} \sim 48^{\circ}$ for quartz. Then
    light emitted in a small azimuth (around the proton direction) $\phi_{ch}$ range arrives directly at the end of the bar; otherwise it
takes a longer path with multiple reflections or, if at larger azimuth angles, is refracted out. About 30\% arrives at the photodetector.
     If several bars are to be read out by a single photodetector, it is important that the light from each bar arrives at the
      photomultiplier at approximately the same time (isochronous design), which occurs with  $\theta_{bar} = \theta_{ch}$ and with the photodetector face
      normal to the bars, as shown in Fig.1. If the bars are to be read out independently, e.g. with SiPMs which have much smaller area than the MCP-PMTs, this is less important. 
      One could then reduce $\theta_{bar}$, but while this increases
      the thickness of bar traversed and the fraction of light collected, it brings the photodetector closer to the beam, which may
      not be allowed by mechanical conflicts, radiation damage and background issues.
      
        \subsection{The L-bar design}

  A novel geometry, the L-bar \textsc{quartic}, combines the virtues of having the Cherenkov radiator bar 
	parallel to the beam (with 100\% of the radiated
      light from protons moving parallel to the bar axis being trapped along the bar) and having the photodetector far 
      from the beam. The bar is L-shaped with a 90$^\circ$ corner. If the surfaces are perfect, no light is refracted out and it all reaches the end of the light 
      guide, except for the light emitted exactly in the plane perpendicular to the LG bar.
     Since $n(\lambda) > \sqrt{2}$ so that
      $\theta_{ch} > 45^\circ$ as it is for quartz,
      the light that passes up the
      LG bar has an angle with respect to the surface that is $< 45^\circ$, less than the critical angle, and total reflection is maintained. 
      This means that the
      path length of the light and number of reflections per unit length are all less than in the radiator bar, 
      which help to allow the
      photodetector to be far from the beam. In addition the blue light path length is less than that of the red light, unlike in the
      radiator bar.
   No mirrors are involved, and the surfaces should not be aluminized as then reflection is not total. All the Cherenkov light, except any that
is absorbed or scattered out by imperfections in the bulk or surface, reaches the photodetector. Another feature of the
L-bar geometry is that it allows segmentation in both $x$ and $y$ directions, which is not the case for the angled-bar solution.

   \begin{figure*}[t]
\centering
\includegraphics[width=70mm]{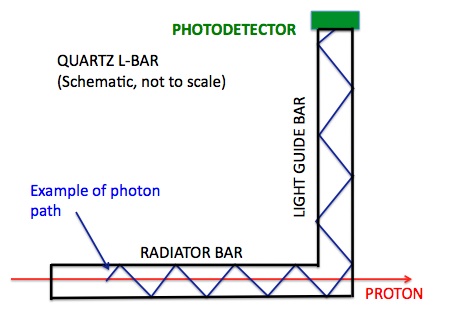}
\caption{Cherenkov light rays in the radiator and light guide bar, for $n$ = 1.48, in the plane of the ``L"
 ($\phi_{ch} = 90^\circ$).}
\end{figure*}

     Consider the section of the L-bar shown in Fig.2, with the radiator bar parallel to the z-axis (beam direction) and the light
     guide bar along the y-axis. For the light rays shown, radiated at angle $\theta_{ch} = cos^{-1}(1/n)$, the speed of
     propagation along the z-axis is:
     \[ \frac{dz}{dt} = \frac{dz}{dr}.\frac{dr}{dt}, \]
     where $r$ is the coordinate along the light path.
     We have $dz/dr$ = cos $\theta_{ch} = 1/n$, and $dr/dt = c/n$, so $dz/dt = c/n^2$.
     The speed of light is, in convenient units, 3 mm/10 ps. So for a L = 20 mm radiator bar the time difference between the light
     emitted at the entrance and at the exit is $\frac{L.(n^2 - 1)}{c}$ = 79.4 ps.
     
     After the light has been trapped in the light guide its angle with respect to the bar axis is the 
     complement, i.e. 90$^\circ
     -\theta_{ch}$, and as the angle in the radiator is $48^\circ$, in the light guide it is $42^\circ$, and it continues to be
      totally reflected. Now
     \[ \frac{dy}{dt} = \frac{dy}{dr}.\frac{dr}{dt}, \]
     and $dy/dr =$ sin$\theta_{ch} = \sqrt{1 - \frac{1}{n^2}}$. 
     Hence the net light speed along the light
     guide bar is:
     \[ \frac{dy}{dt} = \frac{c}{n}\sqrt{1 - \frac{1}{n^2}} = \frac{c}{n^2} \sqrt{n^2 - 1}. \]
     The blue light (larger $n$) now has a shorter path length, partially compensating for its slower speed. Thus
     the dispersion is reduced. The ratio of speeds of propagation along the bars
     $\frac{dz}{dt}/\frac{dy}{dt}$ is tan $\theta_{ch}$, so the effect is modest, about 10\%. The fewer reflections per unit length of the
     radiator bar is another small advantage if TIR is not perfect.

\section{Monte Carlo (\textsc{geant4}) simulations}

We used \textsc{geant4} \cite{geant} to simulate some of the properties if the in-line-, angled- and
L-Bar configurations, to compare with the test beam results, and to aid in the design
of a real HPS detector. Cherenkov photons were generated along the proton path
with wavelength-dependent refractive index $n(\lambda)$ and transmission
$T(\lambda)$ of fused silica. 
The emission polar angle $\theta_{ch}$ is determined by the refractive index $n(\lambda)$, with $cos(\theta_{ch}) = 1/n(\lambda)$, 
and $\phi_{ch}$ is
the azimuthal angle around the beam.
The propagation of the optical photons takes into account the surface and bulk properties
of the bar (wavelength-dependent photon speed and absorption). In the interval $250 < \lambda < 600$ nm about 450 Cherenkov photons
are emitted per cm of radiator.
We calculated the distribution of the time
of arrival of the photons, as a function of $\lambda$, at the MCP-PMT or SiPM. This spectrum was then convoluted with the 
photon detection
efficiency, PDE$(\lambda)$ to simulate the photoelectron time distribution. For each photoelectron a Gaussian time spread 
with width
given by the SPTR was generated and summed, to simulate an output signal. 
We measured the time when the signal  passes 50\% of the signal amplitude (from the proton
arrival time) to emulate a constant fraction discriminator.

The coupling between the bar and and the photodetector is a
potential inefficiency. In the test beam studies grease was used, but it is difficult to simulate correctly.
In the simulations we applied an overall efficiency factor to account for both reflectivity and
coupling losses to match with test beam data.
It is preferable to avoid grease, which can spread to the LG bar; a solid silicon ``cookie" can be used.

The surface reflectivity of the bar is an important factor. In a 100 mm LG bar of 3$\times$3 mm$^2$ cross section
one has typically 30 reflections, so if the average internal reflectivity is only 98\% only 50\% of the light reaches the
photodetector. We aim for close to 100\% internal reflectivity.

\subsection{Simulation of straight bars, in-line or angled}

A bar with the dimensions $3 \times 3 \times 40$ mm$^{3}$ was used to estimate the
timing properties for \textsc{quartic}s with straight bars. We simulated the bar both at $\theta_{bar} = 48^\circ$ to the beam and at 90$^\circ$, 
perpendicular, with the beam central.
In the L-bar case the perpendicular configuration corresponds to particles through the light guides.
 The time spectrum of photons
arriving on the side with the photodetector is shown in Fig.3.
For the perpendicular bar (left plot) the second peak is caused by photons emitted in the
opposite direction and reflected from the far side. This does not occur in the angled bar (right plot), as the oppositely directed photons are refracted
out of the bar.

We simulated the performance of straight bars inclined at angles $\theta_{bar}$ to the protons. In the case $\theta_{bar} = \theta_{ch}$ the
light emitted at azimuth $\phi$ in the direction of bar reaches the photodetector with no reflections; light emitted at azimuth up to
$\Delta \phi = 30^\circ$ from the direction of the bar reaches the detector but with longer path lengths. With 15 cm long bars, we set the proton path  at different bar angles $\theta_{bar}$; as it decreases the amount of light
trapped increases, by a factor $\approx$ 2.5 from 48$^\circ$ to 20$^\circ$. 
Ultimately when $\theta = 0$ all the radiated light is trapped. However the time of arrival of the photons has an increasing
spread, as the path length of the proton along the bar increases. Furthermore for straight bars the photodetector gets closer to
the beam as $\theta_{bar}$ decreases. The L-bar design overcomes this difficulty.

\begin{figure*}[t]
\centering
\includegraphics[width=160mm]{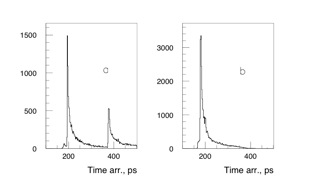}
\caption{The arrival time distributions for (a) perpendicular bar, and (b) bar at Cherenkov angle $\theta_{ch} = 48^\circ$.}
\end{figure*}

\begin{figure*}[t]
\centering
\includegraphics[width=80mm]{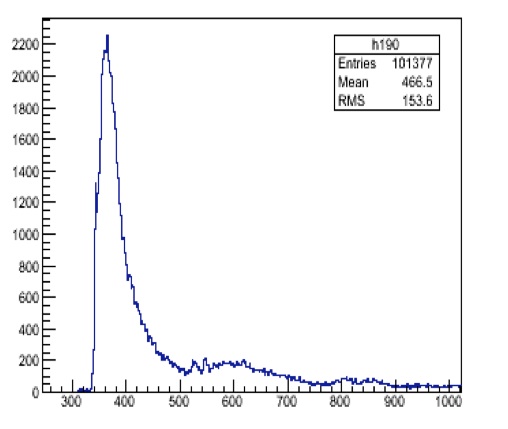}
\caption{\textsc{geant4} simulation showing the arrival time distribution in ps of detected photons (folding in the photon detection
efficiency) at the SiPM for a quartz L-bar, with 30 mm radiator bar length and 100 mm light guide bar.}
\end{figure*}

\subsection{L-bar \textsc{quartic} simulation}

 The simulated photons were propagated by total internal reflection, as well as less-than-total, e.g. 99\% on average per
reflection, to the SiPM. Figure 4 shows the arrival time distribution for photons that produce photoelectrons, from a $3 \times 3 \times 20$  mm$^3$ radiator bar,
with a 40 mm long light guide in the L-bar geometry. The fast leading edge corresponds to photons emitted with azimuth angle
$\phi_{ch} \sim 90^\circ$ close to the LG bar direction (up), and the tail
of that pulse corresponds to other $\phi_{ch}$ angles that have more reflections in the LG. The small later pulse is caused by photons that are not
immediately trapped by the LG bar, are reflected back to the entrance of R and return, where they have another chance of being transmitted up the LG.  
Simulation of radiator bars with both R = 20 mm and R = 40 mm showed the expected shift of the second peak.
Photons with arrival times $t_{PD}$ are detected with the probability given by the photon detection efficiency PDE($\lambda)$, giving the number
of photoelectrons as a function of time. As previously described we simulated the output pulse, taking $\sigma_{SPTR}$ = 38
ps for the MCP-PMT and 120 ps for the SiPM. Then the anode signal distribution for each photoelectron is simulated by a unit area Gaussian with
$\sigma_{anode} =$ 0.5 ns, and these signals added to simulate the output pulse for one proton. We then took the time $t_{meas}$ to be that time 
when the signal passes
10\% (50\%) of its peak value. Fig. 5(a) shows the simulated mean pulse height as a function of radiator bar length, assuming
99\% reflectivity at all surfaces, and Fig.5(b) shows the estimated time resolution. The LG bar lengths are 80 mm. While the
actual resolution depends on some factors that are not well known (e.g. reflectivity and coupling to the photodetector),
the qualitative behaviour of a rapid improvement from R = 10 mm to 25 mm, followed by a slower improvement, is as expected.

\begin{figure*}[t]
\centering
\includegraphics[width=140mm]{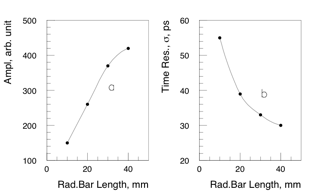}
\caption{a) \textsc{geant4} simulation of the pulse height (arbitrary units) of the L-bar as a function of the radiator bar
length. (b) The simulated time resolution as a function of radiator bar length R. The LG bar is 80 mm and the average reflectivity is 99\%.}
\end{figure*}

The surface reflectivity $R_s$ of the bars is an important factor. A long light guide bar places the SiPM
in a lower radiation field and where it can be better shielded. E.g. with R = 40 mm and LG = 120 mm 100\% TIR gives 1500 photons, while an average
reflection coefficient of $R_s$ = 0.98 gives only 670 (after about 40 reflections; $0.98^{40} = 0.45$).
We found by simulation that with $R_s$ = 0.98 the resolution is $\sigma_t$ = 39 ps (30 ps) for R = 20 mm (40 mm). This is very similar to what was found in the
beam tests. TIR is even more critical with smaller cross section
bars. We will do reflectivity measurements with lasers under different surface conditions to investigate imperfect TIR.

As we expected, the detector simulation shows it to be uniform over the $3 \times 3$ mm$^2$ aperture.  

The full transmission of the Cherenkov light along the radiator bar depends on the perfect matching between the Cherenkov angle 
and the (complement of the) total internal
reflection angle. A small angle between the proton direction and the radiator bar axis will allow some light to leak out. We simulated this; e.g. with a
radiator bar length of 40 mm and an angle of 1/40 = 25 mrad = 1.4$^\circ$ in the plane containing the LG, 20\% of the light is lost. In the HPS application
the accepted protons have an angular spread of $<$ 0.5 mrad, but precise alignment is clearly very important. The directional nature of the L-bar
helps, in this application, to have low sensitivity to backgrounds from other directions.

    \section{Beam tests}
  
  We tested prototype \textsc{quartic}s in the Fermilab test beam (MTest) with 120 GeV/c protons. The trigger counters and detectors under test were
  enclosed in a light-tight and RF shielded (with copper sheet) box~\cite{ronzhin2}. The box had light-tight feed-throughs for high voltage and signal cables.
  A 2$\times$2 mm$^2$ scintillation counter at the front, viewed by two PMTs in
  coincidence, provided a trigger and defined the beam. At the back of the box a 5$\times$5 cm$^2$ counter with a 7 mm diameter central hole, viewed by two PMTs in
  ``OR", was used as a veto to reject events with upstream interactions. For some of the tests a reference time counter 
  was placed at the back (just
  upstream of the veto counter); this was a PHOTEK PMT240 MCP-PMT directly in the beam. Cherenkov light in the 9 mm thick quartz window
  gave a signal of about 35 p.e. with time resolution $\sigma_t \sim $ 8 ps.
  
  We used two types of data acquisition, DAQ. In Phase 1 (DAQ-1), see Ref.~\cite{ronzhintof}, with three MCP-PMT detectors under test, 
  the signals were sent through a constant fraction discriminator (ORTEC 9327) and to an ADC to monitor the pulse heights and allow
  time-slewing corrections to be made. The discriminated pulse was sent to
  time-to-amplitude
  converters (TAC, ORTEC 566) with a pair of detectors as input, followed by analog-to-digital converters (ADC,
  ORTEC 114) read by the on-line computer. 
  The time resolution of the DAQ was measured~\cite{ronzhintof} with a split PiLas laser signal to be $\sigma_t$(DAQ) $\sim 3$ ps.
  The three time differences between the detectors enabled the resolution of each counter to be unfolded. We removed a few percent of the events in
  the tails of the pulse height distribution due to interactions or pile-up (more than one proton in the same RF bucket), and applied
  data-driven time slewing corrections (a small linear correction, determined by plotting the time vs pulse height of a signal).

  In Phase 2 (DAQ-2), with more channels of SiPM, we used waveform digitizing electronics, DRS4~\cite{ritt,ronzhinwfd}. 
  This is a 5 giga samples per second (5 GSPS) waveform digitizer, thus 
  giving the pulse shape sampled every 200 ps and allowing off-line fits with a parametrization optimized for timing. We had two DRS-4 modules
  (boxes)
  each with four channels. One channel in each box was used for a reference signal from the PMT240 (the output was split with a passive
  splitter), and the other three for SiPM
  signals. These SiPM signals were passed through ORTEC VT120 $\times$20 amplifiers. We previously published~\cite{ronzhinwfd} a study of
  SiPM signals, with a 30 mm long in-line quartz Cherenkov radiator with DRS4 electronics. With Hamamatsu 3 $\times$ 3 mm$^2$ 
  MPPC type S10362-330050C, with 3600 pixels of 50 $\mu$m, we measured a resolution of $\sigma_t$ = 30 ps. 
  Various algorithms were tried on the waveform to optimize the time resolution. The rise time from 10\% to 90\% was about 1 ns, i.e. 5
  samplings. Good results were obtained by a linear fit to the two points before and 
  after the 50\% of pulse-height maximum level, finding
  the time at which that line crossed the 50\% level, and then applying a time-slewing correction. More complete waveform fits were
  found to be not significantly better. We did not attempt to correct for small differences in the time between successive samples. We measured the intrinsic resolution of the DRS4 channels \emph{in situ} by splitting the PMT240 signal and sending it to
  two channels of one box. We found~\cite{ronzhinwfd} $\sigma_t$(DRS4) = 6 ps when the time difference between the two signals is very small, but if a
  delay of 2 ns is put in one channel it rises to $\sigma_t$(DRS4) = 20 ps. 
  
   \subsection{Beam tests of straight bars, in-line or angled}
       
      We made preliminary studies with single bars, which we describe first.
      
                We demonstrated~\cite{ronzhin2}, using quartz bars in-line with a SiPM 
directly at the back (a), that the time resolution improves with increasing bar length from 6 mm to 30 mm. With a Hamamatsu 
MPPC (3$\times$3 mm$^2$) and a
30 mm in-line bar we measured, from the width of the pulse height distribution, about 60 photoelectrons and $\sigma_t$ = 14.5 ps, 
and $\sigma_t$ = 35 ps with a 6 mm radiator bar. 
At least over this range the increased light is more important in improving the resolution than the longer pulse is in worsening it. 
      
      We made a pair of detectors each with a single 6$\times$6 mm$^2 \times$ 80 mm quartz bar coupled to a PMT210 (10 mm 
      diameter photocathode) MCP-PMT with UV transparent optical grease (Dow Corning). We mounted them~\cite{ronzhin2} 
      inclined at $\theta_{bar}$ = 48$^\circ$ on
      opposite sides of the beam. The beam size (2 mm) would contribute about 10 ps spread in the time difference, but cancels out in the time
      sum. A PMT240 directly in the beam provided a reference time signal with a resolution $\sigma(t)$ = 7.7 ps. 
      We used the DAQ-1 (see above) electronics. After rejecting $<$10\% of
      events with high pulse height (due to pile-up) and applying a time walk correction (which improved the resolution by $\sim$10\%),
      we obtained (after unfolding) $\sigma(t)$ = 18 ps for the combined pair.

  With a longer, 150 mm bar inclined at 48$^\circ$ with the PHOTEK PMT210
  we found that, within our measurement uncertainties, the signal was constant
  with the beam traversing the bar at different distances from 20 mm to 80 mm from the photodetector. The time resolution degraded slightly
  when the proton distance from the photodetector increased, by about $\Delta\sigma_t \sim$ 1 ps/10 mm.

With several bars in-line, all inclined at $\theta_{ch}$, and a photodetector normal to the bars as in Fig.1,
  the Cherenkov light emitted at azimuth $\phi_{ch}$
   in the direction of the photodetector arrives at the same time from all the bars. With a segmented anode MCP-PMT~\footnote{The
   Burle-Photonis 85011 MCP-PMT has an 8$\times$8 array of 6 mm$\times$6mm pads. We used this in earlier studies, and it is an option
   for the AFP project.
      This design with a multichannel MCP-PMT has segmentation in $x$ (horizontal) but not in $y$ (vertical), 
      which is a disadvantage compared with our new baseline design, the L-bar.} each bar can be individually read, imposing less
      stringent requirements on the electronics and potentially gaining a factor $\sqrt{N_{bar}}$ in resolution over a single bar. With multiple bars on a
      single anode MCP-PMT one gains the same factor in photoelectron statistics, but the electronics must be more highly
      performant.

 We made a pair of multi-bar detectors with single anode \textsc{photek}~\cite{photek} PMT240 (2-stage, 40 mm diameter) 
    MCP-PMTs, see Figs 6 and 7. Despite its large photocathode area, the PMT240 has an isochronous anode
      design; with a pulsed laser scan over the photocathode we verified that the output time is 
      independent of illuminated position to within 2 ps. Each detector had three rows of five 5 mm$\times$ 5 mm bars in a housing made by
      electro-erosion of an aluminum block. Each bar was in a square hole with rounded corners such that it touched the housing only at the
      corners to maintain TIR, and was optically isolated from the other bars.The bars were pushed against the MCP window with springs, and coupled with
      optical grease. Figs.6 and 7 show the internal arrangement of bars and one
  layout in the beam (``opposite side" configuration).  The beam is ($\Delta x$) 2 mm wide (determined by the 
  trigger counter) which contributes up to 35 ps to the time spread in a detector. In this
  configuration the sum of the times (relative to a reference time) in the two detectors is independent of $x$; on the other hand the
  time difference is a measure of $x$. We did not have precision tracking to study this correlation. 
  Even with the constant fraction discriminators (CFD) we found~\cite{ftdalbrow} a residual correlation between the time difference $\Delta t$ and the pulse heights,
  and we applied a linear ``time slewing" correction. The corrected $\Delta t$ distribution was a good fit to a Gaussian distribution with
  $\sigma(\Delta t)$ = 23.2 ps. (The TDC was calibrated with a delay cable.) Comparison with a third reference counter showed that the two detectors
  had the same resolution, 16.0$\pm$0.3 ps each. Combining the pair (``double \textsc{quartic}") as one would in an experiment then has $\sigma(t)$ =
  23.2 ps/2 = 11.6 ps.
  We verified that the time
  resolution $\sigma_t$ improves with the number of bars $N$ as $1/\sqrt{N}$, as shown in Fig.8, showing that the bars contribute about equally; with
  five bars the number of photoelectrons is $\sim$ 20-25.

   \begin{figure*}[t]
\centering
\includegraphics[width=80mm]{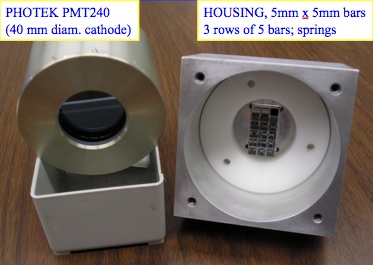}
\caption{The housing made for the PHOTEK PMT240 (left) with three rows of angled bars. The quartz bars are held in a
  ``crate" made by electro-erosion machining, such that the bars are only touched at their corners. Springs at the far ends of the bars
  apply a small pressure on the PMT240 window. Optical grease is used.}
\end{figure*}

\begin{figure*}[t]
\centering
\includegraphics[width=70mm]{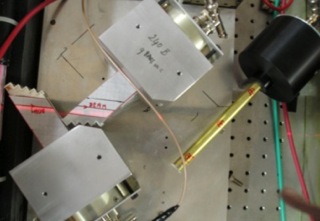}
\caption{Set-up in the beam with PMT240 \textsc{quartic}s on opposite sides of the beam, which comes from the
  left. The brass tube on the right houses the 150 mm long bar on the PMT210.}
\end{figure*}

  \begin{figure*}[t]
\centering
\includegraphics[width=80mm]{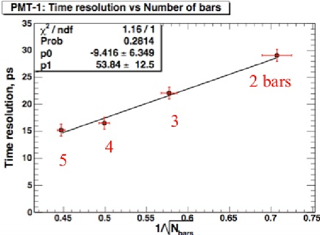}
\caption{The time resolution of angled bars on the PMT240, for different numbers of installed bars, plotted versus
 $1/\sqrt{N}$, showing the expected $\sqrt{N}$ improvement.}
\end{figure*}

\subsection{Beam tests of the L-bar \textsc{quartic}}

We tested prototypes of the L-bar \textsc{quartic} with 120 GeV/c protons in February 2012 (with DAQ-2).  
We measured the signals from four bars in-line in 5 GSPS waveform digitisers (DRS4),
together with the signal from a faster ($\sim$ 8 ps resolution) PHOTEK PMT240 behind the test modules.
 
 We made a pair of identical boxes, each containing two R = 40 mm and two R = 30 mm radiator bars, adjacent to each other, and only
separated by two 100 $\mu$m wires, to maintain total internal reflection; there was no wrapping. Bar holders were made that touched the bars
only at their corners and only at a few positions.  
Fig.9 shows the design of an assembled box, with two 30 mm and
two 40 mm bars in line.

  \begin{figure*}[t]
\centering
\includegraphics[width=80mm]{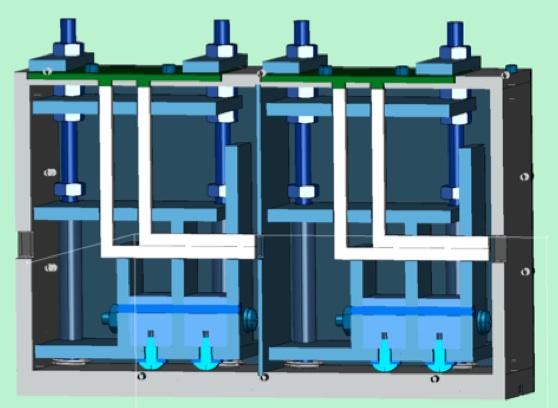}
\caption{Design of an L-bar box with two modules, each with two L-bars, as used in the tests. The beam comes from the right, and the 
SiPMs are at the top.}
\end{figure*}

\begin{figure*}[t]
\centering
\includegraphics[width=90mm]{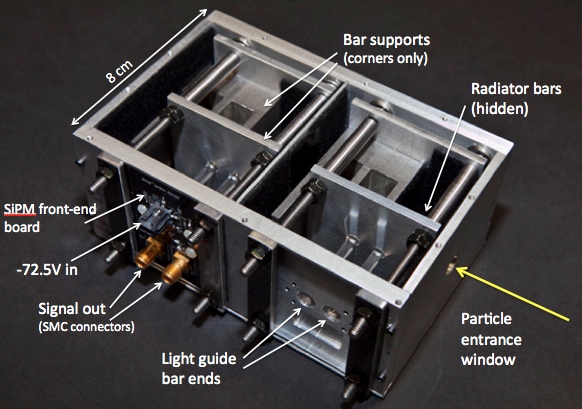}
\caption{Photograph of a test beam box with two sections, each with two L-bars. For one pair the SiPM board is in
place; on the other it is removed to show the bars emerging from their support plate.}
\end{figure*}

Fig.10 shows a photograph.
The interior of the boxes was covered with black felt to absorb any stray light that emerges from the bars. 
The upstream end of the radiator bars could be blackened to
absorb light which is reflected back. We did not do this for these tests, in order to be able to look through both bars in the box (we had small apertures
in the front and back plates) to verify alignment and be able to inspect the bars \emph{in situ}. 
We made two identical boxes, both for convenience and so that they could be
separated in $z$ (the beam direction) as an independent check on the time calibration.

The pair of light guide bars, separated by 10 mm in $z$, were coupled with optical grease
 to Hamamatsu~\cite{hama} 3$\times$3 mm$^2$ SiPMs, MPPC
S10943-0035. These are mounted on a specially designed circuit board, supplied with -72.5 volts, and the signals were output on SMA connectors. 
We took data in two configurations, with and without 9.1 pF shaping capacitors in series. In the former case the clipped signal was put
through an ORTEC VT120 $\times$20 preamplifier. The signals were sent to 8 channels of DRS4 waveform
digitisers~\cite{ronzhinwfd}.

  \begin{figure*}[t]
\centering
\includegraphics[width=70mm]{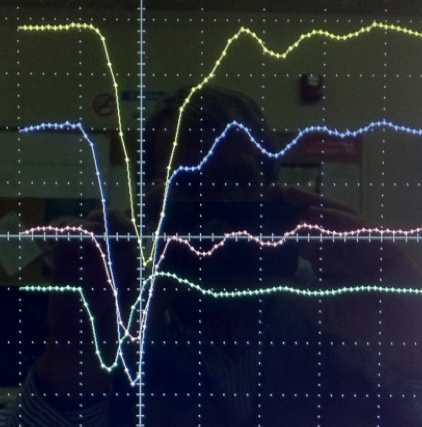}
\caption{Waveforms in the DRS-4 of one proton through three L-bars and the PMT240 (bottom trace). The time scale is 2 ns/div and the vertical
scale 20 mV/div for the top three traces (SiPM's) and 50 mV/div for the bottom trace (reference PMT240). }
\end{figure*}

We sent the 2$\times$2 mm$^2$ beam through four 30 mm (short) radiator bars (with 40 mm light guide bars) and 
      then through four 40 mm (long) bars (43 mm LG bar).
Fig.11 shows a typical event; the green (lowest) trace is the PMT240 signal (50 mV/division), and the other traces are the signals from three 30 mm
bars, with 2 ns/div and 20 mV per division. The pulses are about 80 mV with a rise
time (10\% to 90\%) of $\sim$ 800 ps. From the spread in pulse heights we estimated the number of photoelectrons to be $N_{pe} \sim$ 
80 - 100. We found the signal time from fitting the leading edge (or the full waveform; the difference was small) and correcting for pulse-height
slewing. 
 After these corrections, the time differences between the four short bars and the reference PMT240 signal had widths, from Gaussian
      fits, $\sigma_t$ = 34.9, 39.6, 40.1, and 35.3 ps, so $\langle \sigma_t \rangle$ = 37.5 ps. 
      Fig.12 shows one example, with $\sigma(\Delta $t) = 34.9 ps, and no background or inefficiency.
      The small spread is an indication of only small differences in
      the bars, the SiPMs, or their coupling. 
      We found that the DRS4 resolution depends on the time difference between the input signals~\cite{ronzhinwfd}.
      After unfolding the PMT and electronics resolution (8 ps and $\sim$ 15 ps respectively) 
      we find $\sigma_t$(30 mm bar) $\sim$ 33 ps.
      The time differences between pairs of three different bars in the same DRS4 box were 43.4, 43.9, and 45.2 ps,
      implying a single bar resolution $\sigma_t$(30 mm bar) $\sim$ 31 ps, in agreement.

\begin{figure*}[t]
\centering
\includegraphics[width=80mm]{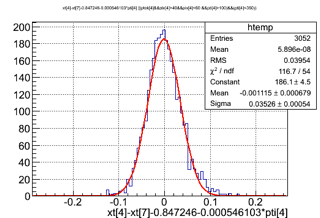}
\caption{The time difference between one L-bar (30 mm radiator, 40 mm light guide, 
  Hamamatsu MPPC type S10362-330050C) and the
  reference time signal (PMT240 in beam).}
\end{figure*}

In the proposed application at the LHC~\cite{fp420} we plan four detectors in line.
Hence the resolution of the four bars combined would be $\sigma_t \approx$ 16 ps.
The  SPTR of the SiPM is quoted by Hamamatsu to be $\sim$ 300 ps;
giving an expected time resolution for 100 p.e. of 300 ps$/\sqrt{100}$ = 30 ps, in reasonable agrement with the observations. SiPMs
with reduced SPTR are becoming available~\cite{stm}.
We checked the time calibration by separating the two detector boxes by 3 cm in the beam direction and observing 
the 100 ps shift between the L-bar
signals. The L-bar geometry has protons that traverse the short bar and cross the light guide of the long bar, and so they 
also have a signal that is about
10\% of the short bar signal, approximately proportional to the proton path length (3 mm). We verified this 
with the proton beam only through the
light guide bars. Protons through the long bar 
did not pass through the short bar light
guides, and we do not see signals there as expected~\footnote{Optical coupling between adjacent bars would show up, but is not seen.}.

We rotated the boxes so that the proton beam passed through the LG bar
at 48$^{\circ}$ as in the angled bar studies, but now with the SiPMs, to compare 
with earlier results using MCP-PMTs. Such angled
bars, with SiPMs on the ends, are an alternative geometry to the L-bar. In this test the lengths of the two bars between the proton and the 
SiPM was only 28 mm and
39 mm. We measured $\sigma_t \sim$ 60 ps in each bar, with about 1/4 the number of photoelectrons. This is much worse that our 
earlier studies of the angled-bar with an MCP-PMT, the reason being that MCP-PMTs have a much better 
SPTR than the SiPMs, and extend further in the UV. (We measured SPTR = 45
ps for the PHOTEK240 MCP-PMT, and 120-150 ps for the Hamamatsu MCCP.)  

Advantages of the L-bar over the angled bar are a 
longer radiator path, for a given distance of the SiPM from the 
beam the light guide bar length
is minimized, and in addition $y$-segmentation is possible.
It would be possible to combine the L-bar geometry with MCP-PMT readoout, if the pattern of anode pads can match that of the bars. If the
photocathode lifetime issue is solved, and cross talk between anode pads\footnote{Unlike in the angled-bar \textsc{quartic}, the signals from
the different bars do not
arrive at the same time.}  is not large, this could be a good option.

 \section{A multichannel L-bar array for the LHC: Design considerations}

\begin{figure*}[t]
\centering
\includegraphics[width=100mm]{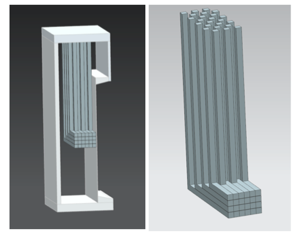}
\caption{Schematic arrangement of a 24-channel module with 3$\times$3 mm$^2$ bars, with a thin wall on the
   LHC beam side (on the right). The SiPM array at the top has space for 3$\times$3 mm$^2$ photodetector areas in 5$\times$5 mm$^2$ packages.
   }
\end{figure*}

   We now present a design for a timing detector using L-bars for the HPS proposal to add very forward tracking and timing 
   detectors to the CMS
   experiment at the LHC~\cite{fp420}. Fig.13 shows the arrangement of bars in a module.
   (A similar design would also be suitable for the APF project at ATLAS.) A ``module" 
   is a light tight box with a very thin ($\sim$ 100 $\mu$m) side wall on the beam side, and blackened interior. The active area is 12 mm
   (vertically, $y$) $\times$ 16 mm (horizontally, $x$), see Fig.13. (This is an example only; other choices of coverage
   are equally possible.) One module consists of (4$\times$6 = 24) independent 3$\times$3 mm$^2$
   bar elements.This allows a time measurement of two or more protons from the same bunch crossing (which has a 
   time spread $\sigma_t \sim$
   150 ps) if they are in different elements. 
   The ends of the light guide bars arrive at an array of SiPMs mounted on a board, together with preamplifiers (and possibly also
   discriminators). A feature of the L-bar design is that the layout of the SiPMs on the 
board (in the ($x,z$) plane) reflects the layout of the bars (in the
($x,y)$ plane) but more spread out in $z$. This is shown in the basic board design in Fig.14.

\begin{figure*}[t]
\centering
\includegraphics[width=100mm]{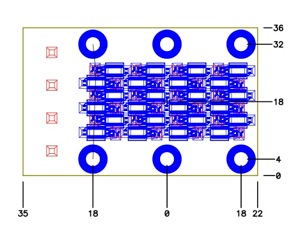}
\caption{Layout of a SiPM board for the 24-channel module shown in Fig. 13. 
   }
\end{figure*}

 The bars can be held in place, but separated to allow TIR, with
   grids of fine wires, diameter about 100 $\mu$m, such that they do not touch and TIR is maintained, in $x$ and $y$. 
   The upsteam end of the bars can be made non-reflecting (e.g. with 
   black felt) to reduce the late 
   bounce-back light. They can also be glued in place at the front without significantly affecting the Cherenkov light. 
   The bars have spring contacts on the SiPM faces, with a thin silicon ``cookie" for good optical coupling.

     The LG can be long enough that the SiPMs can be in shielded enclosures and 
far enough from the beam that radiation doses are tolerable. 
Simulations with \textsc{fluka}~\cite{esposito} show that the hadron flux, with kinetic energy above 50 MeV (from collisions) 
at $y$ = 8 cm above the beam plane is only $\sim 5\times 10^{11}/$cm$^2/100 $ fb$^{-1}$ (one LHC year at high luminosity). A similar 
flux is expected from beam-gas interactions. 
Shielding around the SiPMs can reduce the flux of photons and electrons, as well as low energy neutrons~\footnote{Neutrons with 
energies $\sim$ 1 MeV are
the main concern, and we are making measurements \emph{in situ} during LHC runs. A recent study~\cite{nrad} shows only a few
 percent loss of signal up to 4$\times 10^9 n_{eq}$/cm$^2$, with a rise of dark current which is partially recoverable. Lithiated polyethylene shielding can be installed, if necessary, around the SiPM
boards and even around the LG bars without touching them.}, and only the quartz bars 
and metal housings are very close to the beam.
The main cause of the time spread of the photoelectrons is caused by the length of the radiator bar. The component of the speed of the light along
the radiator bar is $dz/dt = c/n^2 \sim$ 0.136 ($\lambda$ = 300nm) - 0.141 ($\lambda$ = 600 nm) mm/ps.

Independent module positioning gives the option of staggering in $x,y$ for better proton position measurement (to combine with the precision silicon tracking). For example if two
of the four modules are displaced in $x$ by 1.5 mm (half a cell), the \textsc{quartic} measures a 
track with $\sigma(x) \lsim$ 450 $\mu$m, which can help matching to precision silicon pixel or strip track detectors. One
maintains four measurements per track except for the 1.5 mm closest to the beam (but that loss can be recovered with two 
additional cells). Alignment of the radiator bars parallel to the protons is 
critical (at the level of $\lesssim$ 10 mrad). 

An issue with this design is that particles passing through the shorter bars
traverse the light guides of the longer bars at the same $x$, and they will generate light there. This is trapped and
reaches the SiPMs. It is only about 10\% of the light from the traversed bar and should be readily distinguishable (and
one knows the track coordinates). This light can also be used, in the absence of ``spoiling tracks", to 
help with the timing.
Thus for a proton through the shortest bar in Fig.13 we can add the signals from the 3$\times$3 mm$^2$ light guides behind to the
radiator bar signal, reducing the spread in performance between the long and short radiator bars. Another option, if
space allows, is to alternate modules with LG bars above and below the beam pipe (this requires a mirror image
construction), so protons traverse both long bars and short bars.

   These timing detectors will be behind precision silicon tracking, so the hit elements are predetermined, and alignment can be verified. This design
   has flexibity to adapt to needs. For example the highest track density is close to the LHC beam, and one could have smaller bars, e.g. 1
   mm$\times$1 mm, in that area, as well as larger bars in the periphery.

\section{Further developments}

 The L-bar \textsc{quartic} gives time measurements when two or more protons from the same bunch crossing are in its acceptance.
The design of the test beam modules can be extended to 24 channels with only minor developments. 
Nevertheless we will continue R\&D to improve the time resolution.
There are several possibilities. Faster SiPMs with higher photon 
   detection efficiency, and possibly more
   sensitivity in the UV are becoming available~\cite{stm}.
The SiPMs used in the test were 
samples from the CMS Hadron Outer (HO) upgrade, and were not optimised for timing.
SiPMs from STMicroelectronics (STM) with new P-on-N structure (rather than N-on-P SiPMs, also from STM) 
show significantly better timing properties~\cite{mazzillo}~\footnote{We thank STMicroelectronics for providing samples.}. 
Tests with a PiLas (Picosecond
Injection Laser) showed the photon
detection efficiency at $\lambda$ = 405 nm, 5 V above breakdown voltage (28 V), to be  43\% higher (31.1\% cf 21.7\%). Also the 
Single Photoelectron Time Resolution, SPTR, is 174 ps cf 231 ps, i.e. smaller by 25\% than for STM N-on-P detectors. 
Together these improvements lead one to expect that the single bar resolution can be improved from the measured 32 ps to $\sim$20 ps, 
and hence $\sigma_t$ = 10 ps for four modules can be achieved. 

\subsection{Other radiator materials}

   Sapphire (Al$_2$O$_3$) is another potential radiator material, with a higher
refractive index, ($n$ = 1.70 cf 1.47 at 400 nm) resulting in more Cherenkov photons. The
transmission extends down to 250 nm, and sapphire is as radiation hard as quartz. However the time spread over a 
bar of given radiator length goes like $n^2/c$,
and the dispersion is higher. 
\textsc{geant4} simulations (Section 4) are encouraging, and show a factor $\sim$1.9 in the number of detected photons 
with a 30\% improvement in the time resolution.  It is therefore a promising
alternative to quartz; but laboratory measurements need to be done. There may be even better materials.

\subsection{L-bar \textsc{quartic} with MCP-PMT readout}  

It should be possible to replace the SiPM board, with its 24 independent SiPMs, with a custom designed MCP-PMT, 40 mm $\times$ 40 mm, with a single
photocathode and MCP plates, but with a segmented anode with 24 independent pads. The smaller SPTR and better PDE in the blue/UV should improve the time resolution
by a large factor. The Argonne-Chicago-Fermilab team~\cite{frisch} have recently made 20 cm $\times$ 20 cm MCP plates with the anode divided into
strips. Smaller devices with anode pads to match the bars could be developed. If the lifetime, in the harsh LHC conditions, is acceptable, this would be
a very interesting development.

\subsection{Other comments}
If there is longitudinal space one could have more than four modules, gaining as 1/$\sqrt{N}$, as they are
independent, apart from a common reference time signal which has a negligible jitter.
The radiator length was not optimised. The \textsc{geant4} studies show that the resolution improves almost as $\sigma_t \propto 1/R$ over 10 mm $< R < $ 40 mm,
simply because of the increase in total light (almost, but not exactly). So in the same total detector length one could choose
(say) four bars of 40 mm or eight of 20 mm, and get similar resolutions. The 8-bar option would double the cost for little gain, and is not our baseline. 
The length of the LG bar has to be determined based on more radiation studies (including dosimeter
measurements in the tunnel), and on perfecting TIR on the bar sides. The timing algorithm may be improved, and adapted to the read-out, likely
to be an HPTDC (High Precision TDC)~\cite{hptdc} with 25 ps resolution.  

Smaller area SiPMs, with smaller capacitance, are faster; the L-bar design allows smaller bars in the high
density region close to the beam pipe, with larger bars further away. 

A set of L-bar modules can provide a prompt signal for triggering events, at level 1 in CMS for detectors at $z = \pm$ 240 m. The
simplest is to make an ``OR" of the signals from all the bars in a module, then require, for example, at least three out of four modules to have a
signal with fast majority logic, and send a single bit per side to the level 1 trigger. One could also make a more sophisticated
fast track trigger, utilizing the spatial information provided by the individual bars. The individual bar bits, after discriminators, are available
$\sim$ 10 ns after the passage of the proton. The $x$-coordinate information from each module can be input to a look-up table
that can output an approximate proton momentum. This would not be competitive with the silicon tracking, but any reduction of level 1
trigger rates can be important. Time information from the two far detectors can also in principle be used in a fast trigger.

\section{Beam monitoring}

The L-bar \textsc{quartic} design lends itself naturally to the problem of measuring fluxes of particles near (e.g. a few mm away from) high intensity 
beams, either in a colliding beam situation (LHC or CLIC) or at an external beam. It can be directional (the end of the bar far from the
LG should be made non-reflecting), fast, and can have the SiPM (or another photodetector) remote from the beam and in a 
shielded enclosure. For a thesis on some relevant studies see Ref.~\cite{orfanelli}. The detector is designed for 
single proton detection, but for an intense beam it may be more appropriate to have a gas radiator with a thin mirror at the back to a
remote SiPM.

\section{Summary}

We have developed Cherenkov counters using quartz bar radiators and both MCP-PMT and SiPM readout, 
designed to measure the time of protons at the LHC very close to the beam, with resolution
$\sigma_t \sim$ 10 ps. The area required is only $\sim 4$ cm$^2$.
Our latest design, with a novel L-bar geometry, has $\sigma_t$ = 16 ps, with a path for
improvement, and satisfies the other requirements of edgelessness (within about 100 $\mu$m), sufficient radiation hardness, ability to measure
several protons within a bunch (time spread $\sigma$ = 150 ps) and to be active every 25 ns (the bunch separation). In this design each proton
is measured four times for cross-checks and to improve the resolution. The L-bar \textsc{quartic} detectors can also be used to measure, with good time
resolution and directionality, the halo of circulating beams (either inside the vacuum pipe of outside).

\section{Acknowledgements}

Some of the earlier studies were carried out in collaboration with J.Va'vra (LBNL), A.Brandt (University of Texas at Arlington),
J.Pinfold and Shengli Liu (Univ. Alberta). S.Hentschel (Fermilab) designed the L-bar detectors. We thank Jon Howarth (PHOTEK) for 
loans of MCP-PMTs, and
STMicroelectronics for SiPMs. M. Tobin, C. Nicholson and E.A.Wilson (Fermilab summer students) developed \textsc{geant} 
simulations. We thank Aria Soha for test beam support. We thank Chien-Min Kao for support. We thank the U.S. Department of Energy for support through Fermilab.

\end{document}